\pdfoutput=1
\documentclass[acmlarge]{acmart}
\renewcommand\footnotetextcopyrightpermission[1]{}
\fancypagestyle{firstpagestyle}{\fancyhf{}}
\usepackage[utf8]{inputenc}
\usepackage{booktabs}
\usepackage{array}
\usepackage{multirow}
\usepackage[ruled]{algorithm2e}
\usepackage{graphicx}

\begin{document}

\title{Exposing Non-Atomic Methods of Concurrent Objects}

\author{Michael Emmi}
\affiliation{
  \institution{Nokia Bell Labs}
  \country{USA}
}
\email{michael.emmi@nokia.com}

\author{Constantin Enea}
\affiliation{
  \institution{Université Paris Diderot}
  \country{France}
}
\email{cenea@liafa.univ-paris-diderot.fr}


\begin{abstract}

  Multithreaded software is typically built with specialized “concurrent
  objects” like atomic integers, queues, and maps. These objects’ methods are
  designed to behave according to certain consistency criteria like atomicity,
  despite being optimized to avoid blocking and exploit parallelism, e.g.,~by
  using atomic machine instructions like compare and exchange ({\sc cmpxchg}).
  Exposing atomicity violations is important since they generally lead to
  elusive bugs that are difficult to identify, reproduce, and ultimately repair.

  In this work we expose atomicity violations in concurrent object
  implementations from the most widely-used software development kit: The Java
  Development Kit ({\sc jdk}). We witness atomicity violations via simple test
  harnesses containing few concurrent method invocations. While stress testing
  is effective at exposing violations given catalytic test harnesses and
  lightweight means of falsifying atomicity, divining effectual catalysts can be
  difficult, and atomicity checks are generally cumbersome. We overcome these
  problems by automating test-harness search, and establishing atomicity via
  membership in precomputed sets of acceptable return-value outcomes. Our
  approach enables testing millions of executions of each harness each second
  (per processor core). This scale is important since atomicity violations are
  observed in very few executions (tens—hundreds out of millions) of very few
  harnesses (one out of hundreds–thousands). Our implementation is open source
  and publicly available.

\end{abstract}



\maketitle

\section{Introduction}
\label{sec:intro}

Modern computer software is increasingly concurrent. Interactive applications
and services necessitate reactive asynchronous operations to handle requests
immediately as they happen, rather than waiting for long-running operations to
complete. Furthermore, as processor manufacturers approach clock-speed limits,
performance improvements are more-often achieved by parallelizing operations
across multiple processor cores.

However, building concurrent software is notoriously difficult. Besides
coordinating the order of sequenced operations, programmers must also anticipate
all possible ways in which the intermediate steps of concurrent operations can
interfere with one another. The many possible timings in which operations may
access shared memory leads to many possible, and potentially unexpected,
observable program behaviors. Programmers must manage the interaction between
concurrent operations to exclude timings which lead to program errors.
Traditionally this meant manual and error-prone synchronization using operating
system primitives like locks.

Modern software development kits ({\sc sdk}s) such as the Java Development Kit
({\sc jdk}) simplify concurrent programming by providing “concurrent objects”
which encapsulate shared-memory accesses into higher-level abstract data types.
For instance, The {\sc jdk} provides $16$ atomic primitive register types,
e.g.,~with atomic increment methods, and $14$ concurrent data structures,
e.g.,~with atomic offer, peek, and poll methods. Having been designed and
implemented by experts, and scrutinized by a large community of {\sc jdk} users,
these concurrent objects offer high performance and reliability.

Consequently, given the potentially-enormous amount of software that relies on
these concurrent objects, it is important to maintain precise specifications and
ensure that implementations adhere to their specifications. Many methods are
expected to behave atomically, meaning that the results of concurrently-executed
invocations match the results of some serial execution of those same
invocations; they are expected to behave atomically despite the heavy
optimizations employed to avoid blocking and exploit parallelism, e.g.,~by
preferential use of atomic machine instructions like compare and exchange ({\sc
cmpxchg}) over lock-based synchronization; some methods, notably the iterator
methods of {\sc jdk} concurrent data structures, adhere to weaker consistency
criteria. Regardless, identifying expected consistency criteria and exposing
violations is vital, since violations generally lead to elusive bugs that are
notoriously difficult to identify, reproduce, and ultimately repair.

In this work we focus on a single consistency criterion, atomicity, and
demonstrate that a great number of {\sc jdk} concurrent data structure methods
do not behave atomically. In particular, we identify $10$ classes in the {\tt
java.util.concurrent} package which implement queues, deques, sets, and
key-value maps. For each class we select a small set of core methods which are
believed to behave atomically. These core methods represent the most basic
operations, e.g.,~a key-value map’s put, get, remove, and containsKey methods.
We exhibit non-atomic behaviors involving core method invocations with one
non-core method invocation, thus assigning the blame of non-atomicity to the
non-core method. For instance, a behavior in which a key-value map’s size method
returns $2$ when executed concurrently with the sequence {\small\tt [put(0,1);
remove(0); put(2,3)]} would not be atomic, since at no moment are $2$ key-value
mappings stored.

While atomicity violations can signal the inconsistency of one single operation,
e.g.,~when read-only methods like “size” erroneously read transient object
state, they can also signal graver inconsistencies such as broken object
invariants, e.g.,~when mutator methods erroneously write to transient object
state. For instance, in the {\sc jdk}’s ConcurrentSkipListMap, it is possible
for the final invocation in the sequence {\small\tt [clear(); put(1,1);
containsKey(1)]} to return false when the invocation of clear is concurrent with
another put invocation. This indicates a rather insidious breakage of the
object’s representation invariant\footnote{This bug has been independently
identified and fixed recently:
\url{https://bugs.openjdk.java.net/browse/JDK-8166507}}
since the put invocation for the missing
key~$1$ clearly happens after the offending clear invocation.

As both a means of discovering atomicity violations and enabling reproduction,
we exhibit small test harnesses through which each violation is exposed. For
instance, the harness listed in Figure~\ref{fig:harness} exposes the
priorly-discussed atomicity violation in the clear method. Having indexed the
method invocations, the harness records the return value of each invocation and
checks whether the outcome of a given execution, i.e.,~the contents of the
{\small\tt Result} array, corresponds to that of some atomic behavior. The
executions of a given harness are produced by a test automation framework which
may exhibit any concurrent execution of the harness’s sequence methods, e.g.,~by
executing {\small\tt Sequence1} and {\small\tt Sequence2} on separate threads.
In stress testing with this particular harness we witness two possible outcomes:
{\small\tt[null,()\footnote{We write {\tt()} to represent the results of methods
with void return type.},null,true]} and {\small\tt[null,(),null,false]}. Since
the containsKey invocation returns true in every serial execution of these four
invocations, only the first value corresponds to an atomic behavior; the second
value exhibits an atomicity violation. In our experience, a standard
stress-testing framework exhibits this violation approximately once every few
thousand executions.

\begin{figure}
  \begin{centering}
    \begin{minipage}{0.49\linewidth}
      \footnotesize
      \begin{centering}
        \begin{verbatim}
import java.util.concurrent.ConcurrentSkipListMap;

class TestHarness {
    ConcurrentSkipListMap M = new ConcurrentSkipListMap();
    Object[] Result = new Object[4];
    Object VOID = null;

    void Sequence1() {
        Result[0] = M.put(0,0);
    }

    void Sequence2() {
        Result[1] = VOID; M.clear();
        Result[2] = M.put(1,1);
        Result[3] = M.containsKey(1);
    }

    boolean checkResult() {
        return Result[0] == null && Result[1] == VOID
            && Result[2] == null && Result[3] == true;
    }
}
        \end{verbatim}
      \end{centering}
      \caption{A test harness with two invocation sequences.}
      \label{fig:harness}
    \end{minipage}
    \hfill
    \begin{minipage}{0.49\linewidth}
      \begin{centering}

        \begin{tabular}{lcr}
          \toprule
          \multicolumn{3}{c}{ConcurrentSkipListMap: clear} \\
          \multicolumn{3}{c}{\footnotesize\tt [put(0,0)], [clear(); put(1,1); containsKey(1)]} \\
          \midrule
          outcome & atomic? & frequency \\
          \midrule
          {\footnotesize\tt null, (), null, true}  & \checkmark & 2,453,561 \\
          {\footnotesize\tt null, (), null, false} & $\times$   & 289 \\
          \bottomrule
        \end{tabular}

        \caption{Sample result from $1$ second of stress testing.}
        \label{fig:tests:clear}

        \vskip2em

        \begin{tabular}{lcr}
          \toprule
          \multicolumn{3}{c}{ConcurrentSkipListMap: putAll} \\
          \multicolumn{3}{c}{\footnotesize\tt [putAll(\{0=1,1=0\})], [get(0); remove(1)]} \\
          \midrule
          outcome & atomic? & frequency \\
          \midrule
          {\footnotesize\tt (), 1, 0}        & \checkmark & 8,994 \\
          {\footnotesize\tt (), 1, null}     & $\times$   & 9 \\
          {\footnotesize\tt (), null, 0}     & \checkmark & 140 \\
          {\footnotesize\tt (), null, null}  & \checkmark & 1,999,477 \\
          \bottomrule
        \end{tabular}

        \caption{Sample result from $1$ second of stress testing.}
        \label{fig:tests:putAll}

      \end{centering}
    \end{minipage}
  \end{centering}
\end{figure}

Although we find stress testing to be effective at exhibiting non-atomic
outcomes given the right test harness, there are two basic challenges to
completely automating the violation-discovery process. First, writing a test
harness which exposes any given violation is challenging; this generally
requires both insight into the concurrent object’s implementation, as well as an
imagination for how certain concurrent invocations might interfere with each
other given just the right timing. Second, existing methods for checking whether
a given outcome is atomic are cumbersome; they generally require recording the
partial order in which invocations happened, then enumerating the
exponentially-many linearizations of those partially-ordered invocations, while
checking whether some linearization yields the same outcome; the overhead of
enumerating linearizations is prohibitive since it greatly reduces the number of
executions that can be explored for a given harness, and any given outcome may
occur only tens—hundreds of times per million executions.

We address the first challenge, harness generation, by automating the
exploration of candidate test harnesses. To achieve this, we develop a simple
and complete characterization of test harnesses, i.e.,~for which any possible
outcome observed by some program can also be observed by one of our harnesses,
and then enumerate candidate harnesses until finding one which exposes an
atomicity violation. Our characterization is built around the concept of
invocation sequences: each harness is comprised of sequences of invocations of
the given object’s methods; the sequences may be partially ordered, and all but
one of the invocations invoke core methods. For instance, the harness listed in
Figure~\ref{fig:harness} contains two unordered invocation sequences, four
invocations, and two argument values ({\tt 0} and {\tt 1}). Towards finding
small and simple harnesses, we rank harnesses according to their number of
sequences, invocations, and argument values, and explore lower-rank harnesses
before higher-rank harnesses. For example, we would encounter the harness of
Figure~\ref{fig:harness} after having explored harnesses with two sequences,
three invocations, and two values, since their ranks are ordered, and before
having explored harnesses with three sequences, four invocations, and two
values. Of course, this search will only be practical if small harnesses tend to
expose atomicity violations, and if we can exhibit many, e.g.,~millions, of
executions of each harness in a very short time, e.g.,~one second. Our empirical
results demonstrate that this search is practical, depending on a means of fast
and lightweight atomicity checking.

We address the second challenge, lightweight atomicity checking, by specializing
existing approaches~\cite{DBLP:conf/pldi/BurckhardtDMT10} based on
linearizability~\cite{DBLP:journals/toplas/HerlihyW90} to small bounded test
harnesses. During each concurrent execution, these approaches record the partial
order in which an arbitrary number of invocations are executed; at the end of
each execution, atomicity is decided by checking whether at least one
linearization of those invocations yields the same return values as in the original
concurrent execution; atomicity is otherwise violated. While this method is both
sound and complete for atomicity~\cite{DBLP:journals/tcs/FilipovicORY10}, it is
expensive since executions generally have exponentially-many linearizations. To
avoid this cost, we notice that when the number and identities of invocations
are known ahead of time, as is the case in the executions of one single
loop-free test harness, it is possible to avoid enumerating linearizations for
each execution. Instead, we index a given test harness’s invocations, and check
whether a invocation-indexed vector of return values is included in a
pre-computed set of return-value vectors of atomic behaviors. For instance the
{\small\tt Result} array in the test harness of Figure~\ref{fig:harness} stores this
result vector. This approach is advantageous for two reasons: first, the set of
return-value vectors admitted by atomic behaviors can be precomputed once per
test harness, and reused over millions of test executions, each which simply
checks whether their result vector is in this set; and second, since many
linearizations often map to the same result vector, the set of atomic result
vectors is usually much more compact than the set of linearizations. For
instance, there are $4$ possible linearizations of the test harness of
Figure~\ref{fig:harness}, corresponding to the order of {\small\tt Sequence1}’s put
operation in relation to {\small\tt Sequence2}’s operations, yet only $2$ possible
result vectors. This difference grows quickly: for instance, in harnesses with $6$
invocations, we have observed averages of $16$ linearizations to $3$ results.

Empirically, our solutions to these problems are impactful. Our approach enables
testing millions of executions of each harness each second (per processor core).
This amounts to a $30\times$ speedup over the state-of-the-art Line-Up
checker~\cite{DBLP:conf/pldi/BurckhardtDMT10}, which performs systematic
concurrency testing via schedule enumeration, and atomicity checking via
linearization enumeration, spending on average $31.5$ seconds per harness.
Despite our use of stress testing in place of a more systematic enumeration of
executions, our approach exhibits atomicity violations efficiently due to the
sheer number of executions explored per second, and the $30\times$ greater rate
of harness exploration.

In summary we make the following contributions.
\begin{itemize}

  \item We expose atomicity violations in several methods of several {\sc jdk}
  concurrent data structures.

  \item We develop a complete approach for generating test harnesses which
  expose the non-atomicity of one method among core-method invocations.

  \item We develop an approach for the lightweight checking of atomicity which
  is instrumental in the automation of test-harness generation.

  \item We develop a tool which realizes these ideas to automatically derive
  test harnesses exposing each of the reported {\sc jdk} atomicity violations.

\end{itemize}
Our tool implementation is open source and publicly available\footnote{Our tool
is available on GitHub: \url{https://github.com/michael-emmi/violat}} and can be
applied to generate tests for witnessing atomicity violations in any Java
object, so long as that object’s sequential behavior is deterministic, i.e.,~so
that the same invocation sequence always yields the same outcome. Our tool is
lightweight and easy to use, in that it does not require any special
scheduler-perturbing machinery, due to our use of stress testing in place of
schedule enumeration, and does not require user insight, beyond identifying core
methods.

The remainder of this article is organized as follows.
Section~\ref{sec:atomicity} presents the list of atomicity violations we have
discovered in {\sc jdk} concurrent data structures.
Sections~\ref{sec:definitions:constantin:style} and~\ref{sec:testing} describe
our automation for test harness discovery and efficient atomicity checking.
Section~\ref{sec:experiments} reports on empirical results, and
Section~\ref{sec:discussion} discusses the impact and limitations of our
findings. Section~\ref{sec:related} discusses related work, and
Section~\ref{sec:conclusion} gives concluding remarks.

\section{Atomicity in {\sc jdk} Concurrent Data Structures}
\label{sec:atomicity}

In this section we identify atomicity violations across several {\sc jdk}
concurrent data structure classes. These classes have large interfaces
consisting of many diverse methods, likely making it difficult to implement an
entire class atomically. Research in designing concurrent data structures
typically focuses on a small set of \emph{core methods} representing the most
basic operations, e.g.,~a key-value map’s put, get, remove, and containsKey
methods. Additional methods such as putAll and clear are presumably more likely
to exhibit atomicity violations since they are not as heavily studied. While we
are not aware of existing formal proofs stating that the {\sc jdk}’s
implementations of these core methods are actually atomic, we are also not aware
of any atomicity violations among them; our approach assumes their atomicity.

For the remaining non-core methods of a given class, we characterize atomicity
by the return-value outcomes observed in executions with concurrent core-method
invocations. An \emph{outcome} is simply a vector of the invocations’ return
values in a given execution; the indexing of invocations is fixed by a given
test harness. For example, in the sample tests\footnote{We abbreviate test
harnesses by listing their sequences; for instance, the example test harness of
Figure~\ref{fig:harness} is abbreviated in Figure~\ref{fig:tests:clear}.} listed
in Figures~\ref{fig:tests:clear} and~\ref{fig:tests:putAll}, we observe two
outcomes by stress testing the clear method’s harness, and four outcomes by
stress testing the putAll method’s harness. The clear method removes all
mappings stored in a target key-value map, while the putAll method adds the
mappings from its argument key-value map to the target; both methods have void
return type. The put method adds a mapping to the target key-value map and
returns the previously-mapped value for the given key if one exists; otherwise
put returns false. The get method returns the current value mapped from a
given key, or null if no mapping for that key exists, while containsKey returns
true if the target stores a mapping for the given key. The remove method
removes the current mapping for a given key if one exists and returns the
mapped value, and otherwise returns null.

We say that a given outcome is \emph{atomic} when that outcome is observed in
some serial execution of the same invocations. For example, consider the
possible serial executions in Figure~\ref{fig:serializations} of the putAll
method harness of Figure~\ref{fig:tests:putAll}. There are three possible
serializations, given with their corresponding outcomes. While there are four
possible serializations of the clear method’s harness, each serialization shares
the same outcome {\small\tt[null,(),null,true]}. The observed outcome
{\small\tt[null,(),null,false]}, in which containsKey returns false after the
corresponding key was added, is thus non-atomic. This notion of atomicity
captures the intuitive notion that the user of a given concurrent object would
not be able to distinguish it from an inefficient reference implementation in
which invocations were completely synchronized, and would not execute
concurrently.

\begin{figure}
  \centering
  \footnotesize
  \begin{tabular}{llll}
    \toprule
     \multicolumn{1}{c}{serialization of} & \multicolumn{3}{c}{outcome} \\
    \cmidrule(lr){1-1}  \cmidrule(lr){2-4}
    \small\tt [putall(\{0=1,1=0\})], [get(0); remove(1)] & \tt putAll & \tt get & \tt remove \\
    \midrule
    \small\tt [putall(\{0=1,1=0\}); get(0); remove(1)] & \small\tt (), & \small\tt 1, & \small\tt 0  \\
    \addlinespace
    \small\tt [get(0); putall(\{0=1,1=0\}); remove(1)] & \small\tt (), & \small\tt null, & \small\tt 0 \\
    \addlinespace
    \small\tt [get(0); remove(1); putall(\{0=1,1=0\})] & \small\tt (), & \small\tt null, & \small\tt null \\
    \bottomrule
  \end{tabular}
  \caption{The serializations of the test harness of Figure~\ref{fig:tests:putAll}.}
  \label{fig:serializations}
\end{figure}

Figures~\ref{fig:tests:clear} and~\ref{fig:tests:putAll} depict actual sample
results of stress testing the given test harnesses over millions of executions
in one single second, with observed outcomes collected. As these results
demonstrate, atomicity violations may manifest rarely, which can be problematic
when attempting to diagnosing anomalistic behavior. See
Section~\ref{sec:experiments} for details about the stress testing we perform.
While the atomicity violation in the putAll method signals an inconsistency in
one single operation, violations can have much broader consequences. For
instance, the violation exhibited by the clear method indicates a lasting
inconsistency for possibly all subsequent operations, since the invocation to
containsKey returns false even after its argument key was added subsequently to
clearing.

The operations of {\sc jdk} concurrent data structures are broadly
characterized according to the following categories; Figure~\ref{fig:violations}
list an example violation in each category.
\begin{itemize}

  \item[R-1] Read-only snapshot operations like toString, toArray, keySet, and
  size, which return a complete view of the entire data structure.

  \item[R-2] Read-only snapshot predicates like isEmpty, which return a reduced
  view of the entire data structure.

  \item[R-3] Read-only query operations like contains, containsAll,
  containsValue, get, and peek, which return a reduced view including only
  some elements of the data structure.

  \item[M-1] Mutating bulk operations like putAll, addAll, removeAll, which
  add or remove multiple elements.

  \item[M-2] Mutating reset operations like clear, which to some predefined
  state, e.g.,~the empty state.

  \item[M-3] Mutating local operations like offer, poll, put, and remove, that
  add or remove a single element.

\end{itemize}
Atomicity violations in the read-only categories are generally due to the
offending operation performing reads from multiple distinct transient states;
violations in the mutator categories generally make some transient state visible
to other operations, or interfere with other operations by writing to transient
state. Intuitively, the mutating operations present a greater problem since they
are potentially capable of breaking their given object’s representation
invariant, thus compromising the integrity of all future operations.

\begin{figure}
  \begin{minipage}{\linewidth}
    \begin{tabular}{lllc}
      \toprule
      class & harness & outcome & category \\
      \midrule
      \small ConcurrentLinkedQueue
      & {\footnotesize\tt [offer(1); poll(); offer(0)], [{\bf toArray}()]}
      & {\footnotesize\tt true, 0, true, [1,0]}
      & R-1
      \\
      \addlinespace
      \small ConcurrentHashMap
      & {\footnotesize\tt [containsKey(1); {\bf isEmpty}()], [put(1,0)]}
      & {\footnotesize\tt true, true, null}
      & R-2
      \\
      \addlinespace
      \small ConcurrentLinkedQueue
      & {\footnotesize\tt [poll(); offer(1)], [offer(0); {\bf containsAll}(\{0,1\})]}
      & {\footnotesize\tt 0, true, true, true}
      & R-3
      \\
      \addlinespace
      \small ConcurrentLinkedQueue
      & {\footnotesize\tt [offer(1); poll()], [{\bf removeAll}(\{1,1\})]}
      & {\footnotesize\tt true, 1, true}
      & M-1
      \\
      \addlinespace
      \small ConcurrentLinkedQueue
      & {\footnotesize\tt [poll()], [offer(1); offer(0); {\bf clear}()]}
      & {\footnotesize\tt 0, true, true, ()}
      & M-2
      \\
      \addlinespace
      \small ConcurrentLinkedDeque
      & {\footnotesize\tt [poll()], [{\bf offerFirst}(0); offer(1)]}
      & {\footnotesize\tt 1, true, true}
      & M-3
      \\
      \bottomrule
    \end{tabular}
    \caption{Atomicity violations in {\sc jdk} methods. In the case of ConcurrentLinkedDeque, {\tt offer} and {\tt poll} have a FIFO semantics adding and removing a value at the end and respectively, from the beginning of the queue.}
    \label{fig:violations}
  \end{minipage}
\end{figure}

%
\begin{figure}
  \renewcommand{\thefootnote}{\fnsymbol{footnote}}
  \newcolumntype{P}[1]{>{\raggedright\arraybackslash}p{#1}}
\begin{tabular}{P{0.12\linewidth}P{0.4\linewidth}P{0.4\linewidth}}
  \toprule
  \multicolumn{3}{c}{ConcurrentHashMap} \\
  \cmidrule{1-3}
  core & atomic (?) & non-atomic \\
  \cmidrule{1-3}
    put, get, remove, containsKey
  & putIfAbsent, replace
  & putAll, clear, contains, containsValue, isEmpty, elements, entrySet,
    keys, keySet, values, size, mappingCount, toString
  \\
  \addlinespace
  \addlinespace
  \midrule
  \multicolumn{3}{c}{ConcurrentSkipListMap} \\
  \midrule
    put, get, remove, containsKey
  & putIfAbsent, replace, ceilingKey, floorKey, firstKey, lastKey,
    higherKey, lowerKey, keySet, isEmpty
  & putAll, clear, containsValue, entrySet, values, tailMap,
    headMap\footnotemark[1],
    subMap\footnotemark[1],
    size, toString
  \\
  \addlinespace
  \addlinespace
  \midrule
  \multicolumn{3}{c}{ConcurrentSkipListSet} \\
  \cmidrule{1-3}
    add, remove, contains
  & ceiling, floor, first, last, lower, higher, isEmpty
  & addAll, removeAll, clear, pollFirst, pollLast, containsAll,
    tailSet, headSet, subSet, size, toString, toArray
  \\
  \addlinespace
  \addlinespace
  \midrule
  \multicolumn{3}{c}{ConcurrentLinkedQueue, LinkedTransferQueue} \\
  \cmidrule{1-3}
    offer, peek, poll
  & add, addAll\footnotemark[2], isEmpty, remove, remove(Object), contains,
    element
  & addAll\footnotemark[2], removeAll, retainAll, clear, containsAll, size,
    toString, toArray
  \\
  \addlinespace
  \addlinespace
  \midrule
  \multicolumn{3}{c}{LinkedBlockingQueue} \\
  \cmidrule{1-3}
    offer, peek, poll
  & put, add, take, remove, element
  & addAll, removeAll, retainAll, clear, remove(Object), isEmpty,
    contains, containsAll, size, remainingCapacity, toString, toArray
  \\
  \addlinespace
  \addlinespace
  \midrule
  \multicolumn{3}{c}{ArrayBlockingQueue, PriorityBlockingQueue,
    LinkedBlockingDeque\footnotemark[3]} \\
  \cmidrule{1-3}
    offer, peek, poll
  & put, add, take, remove, element, clear, remove(Object),
    isEmpty, contains, size, remainingCapacity, toString, toArray
  & addAll, removeAll, retainAll, containsAll
  \\
  \addlinespace
  \addlinespace
  \midrule
  \multicolumn{3}{c}{ConcurrentLinkedDeque} \\
  \cmidrule{1-3}
    offer, peek, poll
  & add, addAll, addLast, contains, element, getFirst, isEmpty
    offerLast, peekFirst, pollFirst, remove, remove(Object), removeFirst,
    removeFirstOccurrence
  & addFirst, clear, containsAll, getLast
    offerFirst, peekLast, pollLast, removeAll, removeLast,
    removeLastOccurrence, retainAll, size, toString, toArray
  \\
  \bottomrule
\end{tabular}

  \caption[Non-atomic methods in {\sc jdk} concurrent data structures.]{
    Non-atomic methods in {\sc jdk} concurrent data structures.
    \renewcommand{\thefootnote}{\fnsymbol{footnote}}
    \footnotemark[1]{Non-atomicity of headMap and subMap is inferred from
      non-atomicity in ConcurrentSkipListSet.}
    \footnotemark[2]{The addAll method is non-atomic in LinkedTransferQueue
      and possibly atomic in ConcurrentLinkedQueue.}
    \footnotemark[3]{The LinkedBlockingDeque class provides additional methods.}
  }
  \label{fig:results}
\end{figure}

Figure~\ref{fig:results} lists our findings for the methods of several {\sc jdk}
concurrent data structures. We have selected $10$ of the $14$ concurrent data
structures appearing in the java.util.concurrent package. Those not selected are
the copy-on-write data structures (CopyOnWriteArrayList, CopyOnWriteArraySet)
and DelayQueue for which we have not identified atomicity violations — since
they use coarse-grained locking operations in place of more efficient atomic
instructions like compare and exchange, they are less likely to contain
violations — and the SynchronousQueue, which is not specified using sequential
abstract data types since, e.g.,~since their specifications state that “each
insert operation must wait for a corresponding remove operation by another
thread.” For each class, we list the core methods, followed by the non-core
methods, distinguishing between those for which we do not and do, respectively,
exhibit atomicity violations.

We exclude from our study methods which are not specified using sequential
abstract data types, including:
\begin{itemize}

	\item methods which don't have a sequential semantics, e.g.,~the take method
	of ArrayBlockingQueue which removes the head of the queue, waiting if
	necessary until an element is available,

  \item methods whose semantics involves real-time e.g.,~the overloaded poll
  method of ArrayBlockingQueue which waits up to a specified amount of time if
  necessary for an element to become available, and

	\item methods with nondeterministic return values, e.g.,~the default hashCode
  method implementation inherited from {\small\tt java.lang.Object} whose return
  value is a function of the object’s allocated memory address.

\end{itemize}
For technical simplicity, we have also excluded iteration methods,
e.g.,~iterator and stream, methods with function-valued arguments,
e.g.,~forEach, and methods with mutable output parameters, e.g.,~the
LinkedTransferQueue’s drainTo method.

Overall, we remark that a significant number of method implementations are not
atomic, and that methods with atomic implementations in one class may have
non-atomic implementations in another. According to our classification
exemplified  in Figure~\ref{fig:violations}, we observe the following:
\begin{itemize}

	\item Read-only snapshot methods (R-1) are generally non-atomic, except for
	the implementations in ArrayBlockingQueue, PriorityBlockingQueue, and
	LinkedBlockingDeque.

	\item While the read-only snapshot predicate method (R-2) isEmpty of
	ConcurrentHashMap, LinkedBlockingQueue, and ConcurrentLinkedDeque are
	non-atomic, we suspect that the remaining seven classes’ implementations of
	isEmpty are actually atomic.

	\item The atomicity of read-only query methods (R-3) varies significantly from
	one class to another.

	\item Bulk mutator methods (M-1) are almost always non-atomic, except for the
	addAll methods of ConcurrentLinkedQueue and ConcurrentLinkedDeque.

	\item The clear method (M-2) is nearly always non-atomic, except for in the
	ArrayBlockingQueue, PriorityBlockingQueue, and LinkedBlockingDeque.

	\item The only local mutators (M-3) which we found to be non-atomic were those
	implementing {\sc lifo} semantics in the ConcurrentLinkedDeque, including
	addFirst, pollLast, peekLast, and LinkedBlockingQueue’s overloaded
	remove(Object) method.

\end{itemize}
We discuss possible interpretations for these findings in
Section~\ref{sec:discussion}.

In the following sections we describe our approach to automatically generating
test harnesses. All of the atomicity violations presented in this section have
been exhibited with harnesses generated according to our approach. In
Section~\ref{sec:experiments} we describe an empirical evaluation of this test
harness generation.

\section{Atomicity Against Core Methods}
\label{sec:definitions:constantin:style}

In general, a concurrent object is called atomic (or linearizable) when every individual operation appears to take place instantaneously at some point between its invocation and its return. In order to deal with objects such as those implemented in the {\sc jdk} framework where only some particular operations may break this property, we define a notion of atomicity that relies on a partition of object's methods into core and non-core methods. The core methods are assumed to be atomic while a non-core method is called atomic when there exists no program invoking only that method together with core methods that exhibits a non-atomic behavior. Section~\ref{ssec:harness} defines a notion of \emph{test-harness} which allows to capture any interaction with the concurrent object of any program, and Section~\ref{ssec:atomic_harness} defines the class of \emph{atomic test-harnesses} which admit only atomic behaviors. The definition of atomic non-core methods is formalized in Section~\ref{ssec:atomic_method}.

\subsection{Test-Harnesses}\label{ssec:harness}

Let $\mathbb{V}$ be a set of argument and return values. We define test-harnesses as partially-ordered sets of sequences of method invocations represented concretely as words accepted by the following grammar:
\begin{align*}
invoc & ::=  m(argument^*) \mbox{ where $m$ is a method name and $argument\in\mathbb{V}$} \\
invoc\text{-}seq & ::=  invoc \ |\ invoc\text{-}seq; invoc\text{-}seq \\
list\text{-}invoc\text{-}seq & ::= [invoc\text{-}seq]\ | \ list\text{-}invoc\text{-}seq, list\text{-}invoc\text{-}seq \\ 
happens\text{-}\mathit{before} & ::= i < j \ |\ happens\text{-}\mathit{before}, happens\text{-}\mathit{before} \mbox{ where $i,j\in\mathbb{N}$}\\
harness & ::= list\text{-}invoc\text{-}seq,\{happens\text{-}\mathit{before}\}
\end{align*}
This grammar generates pairs formed of a list of invocation sequences $list\text{-}invoc\text{-}seq$ together with happens-before constraints between those sequences. The happens-before constraints are a list of constraints $i<j$ interpreted as the $i$-th sequence in $list\text{-}invoc\text{-}seq$ happens-before the $j$-th sequence. The following are examples of harnesses generated by this grammar:
\begin{align}
&{\tt[put(0,0)]},\ \ {\tt [clear(); put(1,1); containsKey(1)]} \label{eq:harn1} \\
&{\tt[put(0,0); put(2,0)]},\ \ {\tt [clear(); put(1,1)]} \label{eq:harn2} \\
&{\tt[put(0,0); put(2,0)]},\ \ {\tt [clear(); put(1,1); containsKey(1); get(2)]} \label{eq:harn3} \\
&{\tt[put(0,0); put(2,0)]},\ \ {\tt [clear(); put(1,1); containsKey(1); get(2)]}, \ \ {\tt [put(3,1)]},\ \ \{0 < 2, 1 < 2\} \label{eq:harn4}
\end{align}

A harness corresponds to a program where every sequence of invocations is executed in a different thread. Figure~\ref{fig:harness} gives a concrete implementation of the harness in (\ref{eq:harn1}) where the methods {\tt Sequence1} and {\tt Sequence2} will be executed in two different threads. Happens-before constraints $i<j$ are implemented using for instance condition variables: the sequence $j$ awaits a signal placed at the end of the sequence $i$.

Harnesses represent basic programs that contain no loops or conditionals and have a particular form of synchronization (to implement happens-before constraints). However, they are enough to capture any possible object behavior obtained with an arbitrarily complex program. For any object behavior, i.e., a set of method invocations with their return values and a happens-before relation between these invocations, there exists a harness making exactly that set of invocations which admits an execution with the same happens-before relation between those invocations.

For a harness $h=list\text{-}invoc\text{-}seq,\{happens\text{-}\mathit{before}\}$, we define the following notations:
\begin{align*}
\mathit{methods}(h) & = \mbox{the set of method names used in $list\text{-}invoc\text{-}seq$} \\
\mathit{values}(h) &=  \mbox{the set of different arguments to the invocations in $list\text{-}invoc\text{-}seq$} \\
\#\mathit{sequences}(h) &= \mbox{the length of $list\text{-}invoc\text{-}seq$} \\
\mathit{invocations}(h,m) &= \mbox{the multiset of invocations of the method $m$ in $list\text{-}invoc\text{-}seq$} \\
\mathit{invocations}(h) &= \mbox{the multiset of invocations in $list\text{-}invoc\text{-}seq$} \\
\#\mathit{invocations}(h,m) &= \mbox{the total number of invocations of the method $m$ in $list\text{-}invoc\text{-}seq$} \\
\#\mathit{invocations}(h) &= \mbox{the total number of invocations in $list\text{-}invoc\text{-}seq$}
\end{align*}
For example, in the case of the harness $h$ in (\ref{eq:harn4}), we have that $\mathit{methods}(h)=\{\text{put, clear, containsKey, get}\}$, $\mathit{values}(h)=\{ {\tt 0}, {\tt 1}, {\tt 2}, {\tt 3}\}$, $\#\mathit{sequences}(h)=3$, $\mathit{invocations}(h,\text{put})=\{\{{\tt put(0,0)}, {\tt put(2,0)}, {\tt put(1,1)}, {\tt put(3,1)}\}\}$, and $\#\mathit{invocations}(h)=7$.

We assume that harnesses are words accepted by this grammar that satisfy obvious well-formedness conditions, i.e., the list of arguments in every invocation is consistent with the method signature, and the list of happens-before constraints uses values $i$ and $j$ which are smaller than the length of $list\text{-}invoc\text{-}seq$ and defines a strict partial-order.

\subsection{Atomic Test-Harnesses}\label{ssec:atomic_harness}

Checking whether a harness exposes a non-atomic behavior using testing requires being able to enumerate and check millions of executions of that harness within a small amount of time (to make the enumeration of harnesses practical). Checking atomicity of an execution based on the standard notion of linearizability~\cite{DBLP:journals/toplas/HerlihyW90} is impracticable once test-harnesses contain more than 2-3 invocations. This essentially follows from the fact that linearizability is NP-hard in general~\citep{DBLP:journals/siamcomp/GibbonsK97}.

We thus propose a notion of atomicity for test-harnesses which in general is weaker than linearizability, where by a linearizable harness we mean that all its executions are linearizable. This notion is complete in the limit, i.e., for every violation to linearizability there exists a test-harness which exposes that violation according to our notion of atomicity.

We consider a fixed indexing function $invoc\text{-}index$ which for every harness $h$ fixes an order between the method invocations in $h$, i.e., for every harness $h$,
\begin{align*}
invoc\text{-}index(h): invocations(h) \rightarrow [0,\#invocations(h)-1]
\end{align*}
is a one-to-one correspondence ($invocations(h)$ is the multiset of invocations in $h$). For instance, given any harness $h=(list\text{-}invoc\text{-}seq,happens\text{-}\mathit{before})$, $invoc\text{-}index$ could order invocations as they are listed in $list\text{-}invoc\text{-}seq$. Then, we define an \emph{outcome} of a harness $h$ as a mapping $o$ from invocation indices to return values
\begin{align*}
o: [0,\#invocations(h)-1] \rightarrow \mathbb{V}\cup\{()\}
\end{align*}
which is extracted from an execution of the harness (over the concurrent object under test) as follows: $o(i)=v\in\mathbb{Z}$ iff the invocation indexed by $i$, i.e., $invoc\text{-}index(h)^{-1}(i)$, returns $v$, and $o(i)=()$ when the invocation indexed by $i$ is of a method returning void~\footnote{We consider only executions where all the method invocations terminate.}. Let $\mathit{outcomes}(h)$ be the set of all outcomes of $h$. Examples for the set of outcomes of a given harness are given in Figure~\ref{fig:tests:clear} and Figure~\ref{fig:tests:putAll}.

A \emph{linearization} of a harness $h$ is a sequence of method invocations consistent with the sequences and the happens-before constraints in $h$. Let $\mathit{atomicOutcomes}(h)$ be the set of all outcomes of $h$ extracted from executions corresponding to \emph{linearizations} of $h$. As an example, the set of atomic outcomes for the harness ${\tt [putall(\{0=1,1=0\})], [get(0); remove(1)]}$ can be found at the beginning of Section~\ref{sec:atomicity}.

We say that a harness $h$ is \emph{atomic} when it generates only atomic outcomes, i.e.,
\begin{align*}
\mathit{atomic}(h)=\mathsf{true} \mbox{ iff }outcomes(h) \subseteq atomicOutcomes(h)
\end{align*}

This notion of atomicity is weaker than standard linearizability which compares \emph{histories} instead of outcomes extracted from concurrent and respectively, sequential executions. For a brief recall,
a \emph{history} of a harness $h$ is a triple $hist=(h,o,<)$ where $o$ is an outcome extracted from an execution $e$ and $<$ is the happens-before order between invocations in $e$ (an invocation happens-before another one if it returns before the other one is called)~\footnote{Note that the happens-before order $<$ in the execution $e$ may be stronger than the one included in the harness $h$. For instance, a harness with two parallel sequences (no happens-before constraints) has executions where no two invocations overlap whose happens-before is a total order.}. Informally, a history $hist$ is \emph{linearizable} when it can be linearized to a sequence of invocations that is admitted by the object under consideration. In more precise terms, $hist=(h,o,<)$ is linearizable when there exists another history $hist'=(h,o,<')$ such that $<'$ is a total order and consistent with $<$, i.e., every two invocations ordered by $<$ are ordered in the same way by $<'$. A harness is \emph{linearizable} when all its histories are. Examples of linearizable and non-linearizable histories (for a harness calling methods of a key-value map) are given in Figure~\ref{fig:lin_example}. The linearizable history can be linearized to the sequence $\tt [put(0,0); size(); put(1,1)]$ while the non-linearizable one cannot be linearized to a sequence of the same invocations because {\tt size()} happens after {\tt put(0,0)} while it returns that the map contains no key-value pair (and the only parallel invocation is that of the method put).

\begin{figure}[t]
\begin{center}
  \begin{tabular}{llll}
    \toprule
     \multicolumn{1}{c}{harness} & \multicolumn{3}{c}{atomic outcomes } \\
    \cmidrule(lr){1-1}  \cmidrule(lr){2-4}
    \small\tt [size()], [put(0,0); put(1,1)]  &\hspace{2.7cm}  \small\tt  (0,null,null) & \tt \small\tt  (1,null,null) & \small\tt  (2,null,null) \\[1mm]
    \midrule \\[-3mm]
    Linearizable history: & \hspace{2cm}Harness & \hspace{2mm}Outcome & Happens-before \\[2mm]
    & \small\tt [size()], [put(0,0); put(1,1)] &\small\tt (1,null,null) & \small\tt put(0,0) < size() \\
    &&& \small\tt put(0,0) < put(1,1) \\[2mm]
    \midrule
    Non-linearizable history: & \hspace{2cm}Harness & \hspace{2mm}Outcome & Happens-before \\[2mm]
    & \small\tt [size()], [put(0,0); put(1,1)] &\small\tt (0,null,null) & \small\tt put(0,0) < size() \\
    &&& \small\tt put(0,0) < put(1,1) \\
    \bottomrule
  \end{tabular}
\end{center}
\caption{Examples of (non-)linearizable histories.}
\label{fig:lin_example}
\end{figure}

\begin{proposition}\label{prop:lin}
A linearizable harness is atomic.
\end{proposition}
\begin{proof}
Let $o$ be an outcome of a harness $h$ extracted from an execution $e$, and $hist=(h,o,<)$ the history of $e$. By hypothesis, there exists a history $hist'=(h,o,<')$ of $h$ where $<'$ is a total order consistent with $<$. This history is extracted from an execution $e'$ which corresponds to a linearization of $h$ and thus, $o\in atomicOutcomes(h)$.
\end{proof}


Note that the reverse of Proposition~\ref{prop:lin} is not true. A harness where every non-linearizable history has an atomic outcome is atomic. For instance, the harness in Figure~\ref{fig:lin_example} may be atomic if the only non-linearizable history is that given in the same figure (the outcome of this history is atomic).

However, the following result shows that checking atomicity of harnesses allows to discover all the linearizability violations in the limit.

\begin{theorem}\label{th:lin-atom}
For every non-linearizable harness $h$, there exists a non-atomic harness $h'$ with $\mathit{invocations}(h)=\mathit{invocations}(h')$.
\end{theorem}
\begin{proof}
Let $h$ be a non-linearizable harness. Then, there exists a history $hist=(h,o,<)$ which is not linearizable. We define a harness $h'$ where every invocation sequence is a singleton, containing one of the invocations in $h$, and the happens-before constraints encode the order $<$ from $hist$. Assuming by contradiction that $h'$ is atomic contradicts the hypothesis that $hist$ is not linearizable.
\end{proof}

\subsection{Atomic Methods}\label{ssec:atomic_method}

Based on the notions of harness and atomic harness, we define atomicity against core methods as follows:

\begin{definition}\label{def:atomic_meth}
For a set of core methods $Core$, a method $m$ is called \emph{atomic against $Core$} when
\begin{align*}
\mbox{for every harness $h$ with $\mathit{methods}(h)=Core\cup\{m\}$ and $\#\mathit{invocations}(h,m)=1$, $h$ is atomic}.
\end{align*}
\end{definition}
We omit the reference to $Core$ when it is understood from the context. By Theorem~\ref{th:lin-atom}, replacing ``$h$ is atomic'' with ``$h$ is linearizable'' in Definition~\ref{def:atomic_meth} results in a notion of atomicity which is equivalent to the original one. Section~\ref{sec:atomicity} shows examples of non-atomic methods in various {\sc jdk} concurrent objects.

\section{Checking Non-Atomicity Against Core Methods}
\label{sec:testing}

We present an automatic approach for checking method non-atomicity outlined by the abstract algorithm in Algorithm~\ref{alg:abstract}, which is based on an automatic enumeration of test-harnesses. Automating this process is
important for several reasons. First,
atomicity violations may occur very rarely, once in millions of executions of a given test-harness (see Section~\ref{sec:experiments} for our experimental data), and finding  a harness which has a big enough rate of violations to be exposed using testing is impossible without a deep understanding of the implementation, which we want to avoid.
Then, harnesses are very brittle since very small variations of non-atomic harnesses may become atomic.
For instance, on the left of Figure~\ref{fig:variations}, we list 4 harnesses which expose an atomicity violation for the method clear in ConcurrentSkipListMap. These harnesses contain the same sequences of invocations modulo their arguments which oscillate between $0$ and $1$. On the right of Figure~\ref{fig:variations}, we list all the other 12 possible ways of assigning parameter values from the set $\{0,1\}$ which fail to exhibit a non-atomic behavior (using stress testing).
In hindsight, a harness exposing the atomicity violation for clear needs that the key inserted in parallel to clear is strictly smaller than the one inserted afterwards. However, deriving such knowledge would be impossible without understanding the implementation of ConcurrentSkipListMap.

\begin{figure}[t]
\begin{tabular}{lll}
& {\small Harnesses without atomicity violations:} \\
& {\footnotesize\tt [put(0,x)], [clear(); put(0,y); containsKey(0)]} \mbox{\footnotesize with ${\tt x}, {\tt y} \in\{0,1\}$} \\
& {\footnotesize\tt [put(1,x)], [clear(); put(1,y); containsKey(1)]} \mbox{\footnotesize with ${\tt x}, {\tt y} \in\{0,1\}$} \\
{\small Harnesses exposing atomicity violations:} \\
{\footnotesize\tt [put(0,0)], [clear(); put(1,0); containsKey(1)]} & {\footnotesize\tt [put(1,0)], [clear(); put(0,0); containsKey(0)]} \\
{\footnotesize\tt [put(0,0)], [clear(); put(1,0); containsKey(1)]} & {\footnotesize\tt [put(1,0)], [clear(); put(0,1); containsKey(0)]} \\
{\footnotesize\tt [put(0,1)], [clear(); put(1,0); containsKey(1)]} & {\footnotesize\tt [put(1,1)], [clear(); put(0,0); containsKey(0)]} \\
{\footnotesize\tt [put(0,1)], [clear(); put(1,1); containsKey(1)]} & {\footnotesize\tt [put(1,1)], [clear(); put(0,1); containsKey(0)]}
\end{tabular}
\caption{Variations of harnesses exposing atomicity violations for the method {\tt clear} in ConcurrentSkipListMap.}
\label{fig:variations}
\end{figure}

We automate method non-atomicity checking using a prioritization scheme that enumerates harnesses in an increasing order with respect to the number of method invocations, the number of values used as parameters in those invocations, and the number of invocation sequences (or equivalently, the number of different threads used to make those invocations). For each harness $h$, atomicity is checked by pre-computing the set of atomic outcomes and then enumerating all possible outcomes while checking for membership in the set of atomic outcomes. We restrict ourselves to objects whose sequential behaviors are deterministic so that the computation of the set of atomic harnesses terminates (each linearization of a harness produces exactly one outcome and the number of linearizations is fixed for a given harness).
Just to simplify the formalization, we assume that methods' parameters and return values are objects that can be instantiated with integers.


\begin{algorithm}[t]
  \SetAlgoLined
  \SetKwInOut{Input}{Input}
  \Input{A concurrent object with atomic core $\mathit{Core}$, and a method $m$ of this object}
  \KwResult{$\mathsf{NON\text{-}ATOMIC}$ iff $m$ is non-atomic.}
  $(invoc,val,seq)$ $\gets$ $(0,0,0)$ \;
  \While{true}{
       $H$ $\gets$ $constructHarnesses ( \mathit{Core}\cup\{m\}, invoc,val,seq )$\;
       \For{each harness $h\in H$}{
           $atOut$ $\gets$ $\mathit{atomicOutcomes}(h)$\;
           \For{each outcome $out$ of $h$}{
              \uIf{$out\not\in atOut$}{
                 \Return $\mathsf{NON\text{-}ATOMIC}$\;
              }
           }
	}
	$(invoc,val,seq)$ $\gets$ $\mathit{nextParams()}$ \;
  }
\caption{Abstract algorithm for checking non-atomicity of a method.}
  \label{alg:abstract}
\end{algorithm}

Algorithm~\ref{alg:abstract} enumerates increasing assignments for the non-fixed parameters of $constructHarnesses$, which are generated by the function $\mathit{nextParams()}$. The latter can be any function returning triples of natural numbers that is \emph{complete} in the sense that for any tuple $(i,v,s)\in \mathbb{N}^3$ it eventually returns a triple $(i',v',s')$ which is greater than $(i,v,s)$ component-wise. An instance of $\mathit{nextParams()}$ incrementing atomically each parameter, i.e., returning a sequence of triples $(1,1,1)$, $(2,2,2)$, $(3,3,3)$, $\ldots$ would satisfy this assumption. For some fixed set of parameters, the function $constructHarnesses$ returns the following set of harnesses:
\begin{align*}
constructHarnesses ( \mathit{Core}\cup\{m\}, invoc,val,seq)= \{ h\mbox{ a harness}\ \mid\ & \mathit{methods}(h)=\mathit{Core}\cup\{m\}, \\
& \#\mathit{invocations}(h,m)=1, \\
& \#\mathit{invocations}(h)=invoc, \\
& \mathit{values}(h)=[0,val-1], \\
& \#\mathit{sequences}(h)=seq\}
\end{align*}
We implicitly assume that the concurrent objects under test are data-independent in the sense that any behavior with some number of distinct values $val$ can be captured when the values are taken from the interval $[0,val-1]$. The set of harnesses returned by $constructHarnesses$ is clearly finite for any given parameter assignment.

For instance, when checking the atomicity of the method clear of ConcurrentSkipListMap, the output of $constructHarnesses( \{\text{put, remove, get, containsKey, clear}\},4,2,2)$ contains all the harnesses listed in Figure~\ref{fig:variations}.

The definition of $constructHarnesses$ given above supports many optimizations that we implement in our tool. As it is, it returns many redundant harnesses which may have a negative impact on the time to find an atomicity violation. A first obvious optimization is to perform some symmetry reduction. We say that two harness are \emph{symmetric} iff one is obtained from the other by changing the order between invocation sequences and accordingly, applying a renaming on the happens-before constraints. For instance, the following is symmetric to the harness in (\ref{eq:harn4}):
\begin{align*}
&{\tt [put(3,1)]}, \ \ {\tt[put(0,0); put(2,0)]},\ \ {\tt [clear(); put(1,1); containsKey(1); get(2)]},\ \ \{1 < 0, 2 < 0\} \label{eq:harn4}
\end{align*}
The output of the function $constructHarnesses$ is constrained such that it doesn't contain symmetric harnesses.

Besides this optimization which is agnostic to our verification problem, we consider more specific optimizations like excluding harnesses that contain only read-only invocations, for instance, excluding the harness ${\tt [get(0); containsKey(1)]},\ {\tt [get(1); isEmpty()]}$,
and excluding harnesses where the method under test for atomicity is read-only, e.g., isEmpty, and forbidden from executing in parallel with update methods from the atomic core, e.g.,
\begin{align*}
{\tt [get(0); remove(1)]},\ \ {\tt [put(0,1); put(1,1)]},\ \ {\tt [get(1); isEmpty()]},\ \ \{0 < 2, 1 < 2\}.
\end{align*}
The last two optimizations rely on a user provided classification of the object's methods into read-only and updates depending on whether they modify or not the object's internal state. For instance, in the case of ConcurrentSkipListMap we have used the following classification:
\begin{align*}
\text{read-only}\,: &\  \text{get}, \text{containsKey}, \text{ceilingKey}, \text{floorKey}, \text{firstKey}, \text{lastKey}, \text{higherKey}, \text{lowerKey}, \text{keySet}, \\
& \text{isEmpty}, \text{containsValue}, \text{entrySet}, \text{values}, \text{tailMap}, \text{headMap}, \text{subMap}, \text{size}, \text{toString}\\
\text{updates}\,: &\ \text{put}, \text{remove}, \text{putIfAbsent}, \text{replace}, \text{putAll}, \text{clear}
\end{align*}

Although the function $\mathit{nextParams}$ doesn't enumerate all the possible parameter assignments and it may skip some harnesses, we show that Algorithm~\ref{alg:abstract} is complete in the sense that it returns $\mathsf{NON\text{-}ATOMIC}$ whenever the input method is not atomic (soundness follows easily from definitions). Informally, the completeness requirement on the function $\mathit{nextParams}$ ensures that any harness which is not atomic is ``embedded'' in some harness explored by this algorithm which is also not atomic. We say that a harness $h_1$ is a \emph{prefix} of another harness $h_2$, denoted by $h_1\preceq h_2$ when each sequence of $h_1$ is a prefix of a different sequence of $h_2$ and all the sequences of $h_2$ having sequences of $h_1$ as a prefix happen before all the remaining sequences, i.e., there exists a one-to-one mapping $f:[0,\#\mathit{sequences}(h_1)-1] \rightarrow [0,\#\mathit{sequences}(h_2)-1]$ such that the $i$-th sequence of $h_1$ is a prefix of the $f(i)$-th sequence of $h_2$ for each $i$, and the happens-before constraints of $h_2$ contain $i<j$ for each $i\in \mathsf{image}(f)$ and $j\not\in \mathsf{image}(f)$. For instance, the harness in (\ref{eq:harn1}) is a prefix of the harnesses in (\ref{eq:harn3}) and (\ref{eq:harn4}) but, it is not a prefix of the harness in (\ref{eq:harn2}). The following lemma shows that harness atomicity is prefix closed.

\begin{lemma}\label{lem:complete1}
For any two harnesses $h$ and $h'$, if $h$ is atomic and $h\preceq h'$, then $h'$ is also atomic.
\end{lemma}

Given a fixed function $\mathit{nextParams}$, let $Harnesses(nextParams)$ be the set of harnesses enumerated by Algorithm~\ref{alg:abstract} with this function. The following lemma is a direct consequence of the definitions.

\begin{lemma}\label{lem:complete2}
For any complete function $\mathit{nextParams}$ and any harness $h$ with $\mathit{methods}(h)=Core\cup\{m\}$ and $\#\mathit{invocations}(h,m)=1$, there exists a harness $h'$ with $h \preceq h'$ and $h'\in Harnesses(nextParams)$.
\end{lemma}

The completeness of Algorithm~\ref{alg:abstract} follows from the two lemmas above. For any harness $h$ showing that a method $m$ is not atomic, Algorithm~\ref{alg:abstract} explores a harness $h'$ which has $h$ as a prefix (by Lemma~\ref{lem:complete2}) which is also not atomic (by Lemma~\ref{lem:complete1}).

\begin{theorem}

  Algorithm~\ref{alg:abstract} returns $\mathsf{NON\text{-}ATOMIC}$ iff the input method $m$ is not atomic.

\end{theorem}

The next section presents our experimental evaluation of an incomplete version of Algorithm~\ref{alg:abstract} where the loops enumerating harnesses and respectively, outcomes of a harness, execute until a timeout is reached.

\section{Empirical Results}
\label{sec:experiments}

In this section we describe our empirical evaluation of our approach to
the detection of atomicity violations described in the previous sections.

We have implemented our approach to detecting atomicity violations in a test
harness enumeration and checking tool\footnote{Our tool is available on GitHub:
\url{https://github.com/michael-emmi/violat}}. The input to the tool includes a class
specification which determines the set of core methods for a given class. In
addition, the tool takes three parameters which bound the enumeration of
harnesses to a given number of invocations, invocations sequences, and argument
values. For a given non-core method, the tool generates all possible harnesses
according to Section~\ref{sec:testing} for the given parameter values. The tool
subjects each test harness to one second of stress testing with the jcstress
test automation framework\footnote{The Java Concurrency Stress tests (jcstress)
is an experimental harness and a suite of tests to aid the research in the
correctness of concurrency support in the {\sc jvm}, class libraries, and
hardware. \url{http://openjdk.java.net/projects/code-tools/jcstress/}} and
returns the first test harness for which jcstress observes a non-atomic outcome.

Architecturally our tool has four distinct components. First, to separate the
generation of harnesses from the details of the jcstress framework, an
enumeration component generates an abstract representation we call
\emph{schemas}. Intuitively, a schema is simply a list of partially ordered
invocation sequences; this representation is language agnostic. Second, an
annotation component generates the list of possible atomic outcomes for each
schema; this step is language dependent, since we must actually execute the
invocations to determine outcomes; our implementation invokes a small Java
program that enumerates the linearizations of the given schema, and records the
outcome of each executed linearization. Third, a translation component generates
generates the actual jcstress test harness corresponding to each schema. The
final testing component executes jcstress on each harness. In principle,
substituting the schema annotation, schema-to-harness translation, and harness
testing components would enable the application of our tool to other languages
besides Java. Only our annotation component, and jcstress, are currently written
in Java; the remainder of our implementation is written in portable Node.js
Javascript.

There are three basic limitations of our current implementation. First, while
the parameter bounding the number of argument values is sufficient for methods
with simple parametric arguments, e.g.,~accepting an arbitrary object, many
methods of {\sc jdk} concurrent data structures, like addAll, take collections
as arguments. Our current implementation fixes the cardinality of these argument
collections to $2$, and generates all possible size-$2$ collections. Second, to
simplify the enumeration of argument values, we generate only integer-valued
arguments, or collections thereof, ranging from $0$ to one less than the
value-bound parameter. Third, while our harness schemas can represent harnesses
with any partial order among invocation sequences, the jcstress framework is
limited to expressing harnesses in which all sequences are unordered, with
possible one method — the test class’s constructor — executed initially, and one
distinguished method — called an “arbiter” — executed finally, after all
sequences complete. While we have yet to discover violations which require more,
this would limit our expressiveness in principle.

Since the number of harnesses for the given parameter values is usually quite
large, the order in which harnesses are tested can make a significant difference
on the exploration time. Our original enumeration of harnesses tends to group
very similar harnesses, for instance, changing only a single argument value from
one harness to the next. This is often suboptimal, since even if there are a
great number of similar non-atomic harnesses, they may all appear very late in
the enumeration. Our current solution thus shuffles the enumeration after
generating harnesses to ensure a more even distribution. We perform shuffling
with fixed pseudorandom-number-generation seed to ensure reproducibility across
runs of our test-harness generation tool. We speculate that generating harnesses
at random, — again, with a fixed seed — rather than enumerating completely, will
also be an effective exploration strategy, and eliminate the bottleneck imposed
by having to enumerate all harnesses initially before shuffling; we are planning
to evaluate this strategy.

We have applied our tool to $10$ classes from the Java SE Development Kit 8u121
as discussed in Section~\ref{sec:atomicity}. Tables~\ref{fig:experiments:1}
and~\ref{fig:experiments:2} list a detailed sample of results of our
experiments, as executed on a $4$-core $4$GHz Intel Core i7 iMac. Results can
vary slightly across runs of our tool for two reasons. First, we do not control
the order in which jcstress runs test harnesses within each of the $100$-harness
chunks we provide it. Thus, if a single $100$-harness chunk contains multiple
non-atomic harnesses, we report whichever is first tested by jcstress. Second,
jcstress exhibits many of the non-atomic outcomes with very low frequency, and
sometimes even fails completely to exhibit possible non-atomic outcomes. In our
experience, any given atomicity violation can be exhibited from multiple,
possibly similar, harnesses in our enumeration; even if jcstress misses the
violation in one harness, it is more likely to catch it later, in a similar
harness. For instance, in a previous run a non-atomic outcome of the toArray
method of LinkedBlockingQueue was discovered in 1 out of 1,074,530 executions
after only 2,900 harnesses and 1,072s, compared with 32,400 harnesses and
12,179s in the current sample.

\begin{table}
  \scriptsize

\begin{tabular}{ccllrr}
\toprule
\multicolumn{6}{c}{ConcurrentHashMap} \\
\multicolumn{2}{c}{enumeration} & \multicolumn{3}{c}{failure} & \\
\cmidrule(lr){1-2} \cmidrule(lr){3-5}
\#I, \#S, \#V & exp / gen & harness & outcome & frequency & time \\
\cmidrule(lr){1-6}
5, 2, 2 & 100 / 331,008 & \tt [put(1,1); {\bf clear}()],[put(0,1); remove(0); remove(1)] & \tt N,N,N,1 & 72 / 1,077,520 & 61s \\
4, 2, 2 & 1,700 / 24,384 & \tt [put(0,0); remove(1)],[put(1,0); {\bf contains}(0)] & \tt N,0,N,F & 6 / 1,508,770 & 593s \\
5, 2, 2 & 600 / 662,016 & \tt [get(1); {\bf containsValue}(1)],[put(1,1); put(0,1); put(1,0)] & \tt 1,F,N,N,1 & 1 / 3,993,110 & 325s \\
3, 2, 2 & 100 / 408 & \tt [put(0,1); put(1,0)],[{\bf elements}()] & \tt N,N,[0] & 3 / 1,665,650 & 33s \\
3, 2, 2 & 100 / 408 & \tt [put(0,1); put(1,0)],[{\bf entrySet}()] & \tt N,N,[1=0] & 23 / 2,688,890 & 8s \\
3, 2, 2 & 100 / 408 & \tt [containsKey(1); {\bf isEmpty}()],[put(1,0)] & \tt T,T,N & 134 / 6,970,780 & 8s \\
3, 2, 2 & 100 / 408 & \tt [put(0,1); put(1,1)],[{\bf keySet}()] & \tt N,N,[1] & 18 / 5,048,060 & 17s \\
3, 2, 2 & 100 / 408 & \tt [{\bf keys}()],[put(0,1); put(1,1)] & \tt [1],N,N & 13 / 1,721,300 & 16s \\
3, 2, 2 & 100 / 408 & \tt [containsKey(1); {\bf mappingCount}()],[put(1,0)] & \tt T,0,N & 125 / 5,511,280 & 8s \\
3, 2, 2 & 100 / 3,264 & \tt [{\bf putAll}(\{0=0,1=0\})],[remove(0); get(1)] & \tt 0,N & 2 / 2,099,190 & 12s \\
3, 2, 2 & 100 / 408 & \tt [get(1); {\bf size}()],[put(1,1)] & \tt 1,0,N & 4 / 6,533,430 & 8s \\
3, 2, 2 & 100 / 408 & \tt [put(0,1); put(1,1)],[{\bf toString}()] & \tt N,N,{1=1} & 120 / 3,948,560 & 27s \\
3, 2, 2 & 100 / 408 & \tt [put(0,1); put(1,0)],[{\bf values}()] & \tt N,N,[0] & 99 / 2,836,280 & 25s \\
\midrule
\multicolumn{6}{c}{ConcurrentSkipListMap} \\
\cmidrule(lr){1-6}
4, 2, 2 & 300 / 12,192 & \tt [{\bf clear}(); put(1,0); put(1,0)],[put(0,0)] & \tt N,N,N & 1,000 / 5,067,830 & 102s \\
4, 2, 2 & 1,700 / 24,384 & \tt [put(0,0); remove(1)],[put(1,0); {\bf containsValue}(0)] & \tt N,0,N,F & 2 / 3,418,150 & 585s \\
4, 2, 2 & 200 / 12,192 & \tt [put(0,1); {\bf entrySet}()],[put(0,0); put(1,1)] & \tt N,[0=1,1=1],1,N & 148 / 1,271,070 & 52s \\
3, 2, 2 & 100 / 3,264 & \tt [put(0,0); containsKey(1)],[{\bf putAll}(\{0=1,1=1\})] & \tt 1,F & 25 / 1,847,900 & 10s \\
4, 2, 2 & 1,200 / 12,192 & \tt [put(0,1); remove(1)],[put(1,1); {\bf size}()] & \tt N,1,N,0 & 1 / 3,898,810 & 408s \\
5, 2, 2 & 600 / 662,016 & \tt [put(0,1); containsKey(0); put(0,0); put(1,0)],[{\bf tailMap}(0)] & \tt N,T,1,N,{0=1,1=0} & 56 / 2,291,450 & 341s \\
4, 2, 2 & 200 / 12,192 & \tt [put(0,1); put(1,1)],[put(0,0); {\bf toString}()] & \tt 0,N,N,{0=0,1=1} & 8 / 2,269,590 & 48s \\
4, 2, 2 & 200 / 12,192 & \tt [put(0,1); put(1,1)],[put(0,0); {\bf values}()] & \tt 0,N,N,[0,1] & 12 / 2,069,060 & 71s \\
\midrule
\multicolumn{6}{c}{ConcurrentSkipListSet} \\
\cmidrule(lr){1-6}
3, 2, 2 & 100 / 640 & \tt [{\bf addAll}([0,1])],[contains(0); add(1)] & \tt T,T,T & 2,645 / 3,589,490 & 9s \\
4, 2, 2 & 500 / 2,816 & \tt [add(0)],[add(1); {\bf clear}(); add(1)] & \tt T,T,F & 9 / 2,942,870 & 155s \\
4, 2, 2 & 500 / 11,264 & \tt [add(1); {\bf containsAll}([1,0])],[remove(1); add(0)] & \tt T,T,T,T & 360 / 3,055,220 & 164s \\
5, 2, 3 & 9,100 / 155,520 & \tt [remove(0); remove(1)],[add(0); add(1); {\bf headSet}(2)] & \tt T,T,T,T,[0] & 106 / 1,643,840 & 3282s \\
4, 2, 2 & 400 / 2,816 & \tt [add(0); contains(1)],[add(1); poll()] & \tt T,T,T,1 & 2 / 1,621,280 & 122s \\
4, 2, 2 & 200 / 2,816 & \tt [add(1); contains(0)],[add(0); {\bf pollLast}()] & \tt T,T,T,0 & 4 / 2,262,770 & 64s \\
4, 2, 2 & 100 / 11,264 & \tt [{\bf removeAll}([0,0])],[add(0); add(0); contains(0)] & \tt T,T,T,F & 86 / 3,625,490 & 28s \\
3, 2, 2 & 100 / 640 & \tt [remove(0)],[add(0); {\bf retainAll}([1,1])] & \tt T,T,T & 2,659 / 3,380,760 & 12s \\
4, 2, 2 & 2,500 / 2,816 & \tt [add(0); remove(1)],[add(1); {\bf size}()] & \tt T,T,T,0 & 1 / 2,785,410 & 878s \\
4, 2, 4 & 7,600 / 131,072 & \tt [{\bf subSet}(0,3)],[add(1); add(0); add(2)] & \tt [1,2],T,T,T & 836 / 1,970,550 & 2780s \\
5, 2, 2 & 1,000 / 90,368 & \tt [{\bf tailSet}(0)],[add(1); add(0); remove(0); remove(1)] & \tt [0],T,T,T,T & 631 / 2,340,020 & 350s \\
5, 2, 2 & 1,100 / 45,184 & \tt [remove(0); remove(1)],[add(0); add(1); {\bf toArray}()] & \tt T,T,T,T,[0] & 1 / 739,050 & 388s \\
5, 2, 2 & 100 / 45,184 & \tt [add(0); add(1); {\bf toString}()],[remove(0); remove(1)] & \tt T,T,[0],T,T & 23 / 1,843,310 & 36s \\
\midrule
\multicolumn{6}{c}{ConcurrentLinkedQueue} \\
\cmidrule(lr){1-6}
4, 2, 2 & 300 / 894 & \tt [poll()],[offer(1); offer(0); {\bf clear}()] & \tt 0,T,T & 1,289 / 4,209,580 & 84s \\
4, 2, 2 & 500 / 3,576 & \tt [poll(); offer(1)],[offer(0); {\bf containsAll}([0,1])] & \tt 0,T,T,T & 9 / 4,120,090 & 170s \\
3, 2, 2 & 100 / 312 & \tt [offer(1); poll()],[{\bf removeAll}([1,1])] & \tt T,1,T & 964 / 4,665,060 & 10s \\
3, 2, 2 & 100 / 312 & \tt [poll()],[offer(1); {\bf retainAll}([0,0])] & \tt 1,T,T & 8,222 / 5,328,260 & 10s \\
4, 2, 2 & 100 / 894 & \tt [poll(); offer(0)],[offer(1); {\bf size}()] & \tt 1,T,T,2 & 3 / 3,458,050 & 13s \\
4, 2, 2 & 100 / 894 & \tt [{\bf toArray}()],[offer(1); poll(); offer(0)] & \tt [1,0],T,1,T & 23 / 1,340,340 & 8s \\
4, 2, 2 & 100 / 894 & \tt [offer(0); poll(); offer(0)],[{\bf toString}()] & \tt T,0,T,[0,0] & 21 / 3,845,190 & 18s \\
\midrule
\multicolumn{6}{c}{LinkedTransferQueue} \\
\cmidrule(lr){1-6}
3, 2, 2 & 100 / 312 & \tt [poll(); poll()],[{\bf addAll}([1,0])] & \tt 1,N,T & 40 / 4,365,650 & 15s \\
4, 2, 2 & 300 / 894 & \tt [poll()],[offer(1); offer(0); {\bf clear}()] & \tt 0,T,T & 1,058 / 4,283,910 & 105s \\
4, 2, 2 & 500 / 3,576 & \tt [poll(); offer(1)],[offer(0); {\bf containsAll}([0,1])] & \tt 0,T,T,T & 10 / 4,384,580 & 166s \\
3, 2, 2 & 100 / 312 & \tt [offer(0); poll()],[{\bf removeAll}([0,0])] & \tt T,0,T & 21,385 / 4,968,110 & 9s \\
5, 2, 2 & 100 / 37,560 & \tt [poll(); peek(); offer(1)],[offer(0); {\bf retainAll}([1,1])] & \tt 0,N,T,T,T & 38,473 / 3,610,180 & 14s \\
4, 2, 2 & 100 / 894 & \tt [{\bf size}()],[offer(0); poll(); offer(1)] & \tt 2,T,0,T & 13 / 3,475,350 & 21s \\
4, 2, 2 & 100 / 894 & \tt [{\bf toArray}()],[offer(1); poll(); offer(0)] & \tt [1,0],T,1,T & 103 / 322,400 & 22s \\
4, 2, 2 & 100 / 894 & \tt [poll(); offer(1)],[offer(1); {\bf toString}()] & \tt 1,T,T,[1,1] & 55 / 1,867,010 & 17s \\
\bottomrule
\end{tabular}

  \caption{Empirical results, Part 1.}
  \label{fig:experiments:1}
\end{table}

\begin{table}
  \scriptsize
  \begin{tabular}{ccllrr}
\toprule
\multicolumn{6}{c}{LinkedBlockingQueue} \\
\multicolumn{2}{c}{enumeration} & \multicolumn{3}{c}{failure} & \\
\cmidrule(lr){1-2} \cmidrule(lr){3-5}
\#I, \#S, \#V & exp / gen & harness & outcome & frequency & time \\
\cmidrule(lr){1-6}
3, 2, 2 & 100 / 312 & \tt [{\bf addAll}([1,1])],[poll(); peek()] & \tt T,1,N & 537 / 2,683,060 & 9s \\
5, 2, 2 & 1,600 / 9,390 & \tt [offer(1); {\bf clear}(); offer(0)],[peek(); poll()] & \tt T,T,0,N & 1 / 2,153,170 & 557s \\
4, 2, 2 & 100 / 3,576 & \tt [offer(1); {\bf containsAll}([1,0])],[poll(); offer(0)] & \tt T,T,1,T & 107 / 2,428,020 & 22s \\
6, 2, 2 & 1,100 / 186,108 & \tt [offer(0); peek(); peek()],[poll(); offer(1); {\bf contains}(0)] & \tt T,1,N,0,T,F & 1 / 1,756,190 & 406s \\
5, 2, 2 & 1,700 / 9,390 & \tt [poll(); offer(0)],[offer(1); peek(); {\bf isEmpty}()] & \tt 1,T,T,0,T & 1 / 3,137,600 & 604s \\
5, 2, 2 & 1,500 / 9,390 & \tt [poll(); offer(0)],[offer(1); peek(); {\bf remainingCapacity}()] & \tt 1,T,T,0,2147483647 & 1 / 2,259,160 & 532s \\
3, 2, 2 & 100 / 312 & \tt [{\bf removeAll}([0,1])],[offer(0); poll()] & \tt T,T,0 & 245,899 / 2,416,360 & 9s \\
5, 2, 2 & 5,300 / 18,780 & \tt [offer(1); peek(); peek()],[{\bf remove}(1); offer(0)] & \tt T,0,N,T,T & 1 / 2,097,120 & 1917s \\
3, 2, 2 & 100 / 312 & \tt [{\bf retainAll}([0,0])],[offer(1); poll()] & \tt T,T,1 & 181,760 / 2,212,120 & 8s \\
5, 2, 2 & 1,500 / 9,390 & \tt [poll(); offer(0)],[offer(1); peek(); {\bf size}()] & \tt 1,T,T,0,0 & 1 / 2,139,590 & 536s \\
6, 2, 2 & 32,400 / 93,054 & \tt [{\bf toArray}(); poll(); offer(1)],[offer(0); peek(); poll()] & \tt [],0,T,T,1,N & 1 / 565,500 & 12179s \\
6, 2, 2 & 3,500 / 93,054 & \tt [poll(); offer(1)],[offer(0); peek(); peek(); {\bf toString}()] & \tt 0,T,T,1,N,[1] & 1 / 1,459,580 & 1267s \\
\midrule
\multicolumn{6}{c}{ArrayBlockingQueue} \\
\cmidrule(lr){1-6}
3, 2, 2 & 100 / 312 & \tt [{\bf addAll}([0,0])],[poll(); poll()] & \tt T,0,N & 251 / 3,434,740 & 10s \\
4, 2, 2 & 500 / 3,576 & \tt [poll(); offer(1)],[offer(0); {\bf containsAll}([0,1])] & \tt 0,T,T,T & 5,860 / 2,042,060 & 156s \\
3, 2, 2 & 100 / 312 & \tt [offer(0); poll()],[{\bf removeAll}([1,0])] & \tt T,0,T & 26,798 / 1,279,280 & 8s \\
3, 2, 2 & 100 / 312 & \tt [poll()],[offer(1); {\bf retainAll}([0,0])] & \tt 1,T,T & 1,119 / 2,087,100 & 9s \\
\midrule
\multicolumn{6}{c}{PriorityBlockingQueue} \\
\cmidrule(lr){1-6}
3, 2, 2 & 100 / 312 & \tt [peek(); {\bf addAll}([1,0])],[poll()] & \tt N,T,1 & 10,288 / 3,062,920 & 9s \\
4, 2, 2 & 100 / 3,576 & \tt [offer(1); {\bf containsAll}([1,0])],[poll(); offer(0)] & \tt T,T,1,T & 8,655 / 2,916,430 & 26s \\
3, 2, 2 & 100 / 312 & \tt [poll()],[offer(0); {\bf removeAll}([0,0])] & \tt 0,T,T & 5,522 / 2,625,610 & 11s \\
3, 2, 2 & 100 / 312 & \tt [{\bf retainAll}([1,1])],[offer(0); poll()] & \tt T,T,0 & 7,990 / 2,968,360 & 20s \\
\midrule
\multicolumn{6}{c}{LinkedBlockingDeque} \\
\cmidrule(lr){1-6}
3, 2, 2 & 100 / 384 & \tt [{\bf addAll}([1,0])],[poll(); peek()] & \tt T,1,N & 9,832 / 2,981,050 & 9s \\
4, 2, 2 & 600 / 4,096 & \tt [offer(1); {\bf containsAll}([1,0])],[poll(); offer(0)] & \tt T,T,1,T & 4,177 / 2,555,780 & 206s \\
3, 2, 2 & 100 / 384 & \tt [offer(1); {\bf removeAll}([1,0])],[poll()] & \tt T,T,1 & 11,613 / 3,146,040 & 8s \\
3, 2, 2 & 100 / 384 & \tt [{\bf retainAll}([0,0])],[offer(1); poll()] & \tt T,T,1 & 84,882 / 3,893,750 & 18s \\
\midrule
\multicolumn{6}{c}{ConcurrentLinkedDeque} \\
\cmidrule(lr){1-6}
3, 2, 2 & 200 / 156 & \tt [poll()],[{\bf addFirst}(0); offer(1)] & \tt 1,T & 163 / 5,461,590 & 42s \\
4, 2, 2 & 200 / 894 & \tt [poll()],[offer(1); offer(0); {\bf clear}()] & \tt 0,T,T & 985 / 4,391,760 & 55s \\
4, 2, 2 & 100 / 3,576 & \tt [poll(); offer(1)],[offer(0); {\bf containsAll}([0,1])] & \tt 0,T,T,T & 7 / 3,882,720 & 28s \\
4, 2, 2 & 100 / 894 & \tt [offer(1); {\bf getLast}()],[offer(0); poll()] & \tt T,E,T,1 & 18 / 4,236,420 & 28s \\
3, 2, 2 & 200 / 156 & \tt [{\bf offerFirst}(0); offer(1)],[poll()] & \tt T,T,1 & 23 / 4,906,360 & 44s \\
4, 2, 2 & 100 / 894 & \tt [offer(1); {\bf peekLast}()],[offer(0); poll()] & \tt T,N,T,1 & 22 / 3,540,930 & 17s \\
4, 2, 2 & 100 / 894 & \tt [offer(0); {\bf pollLast}()],[offer(0); poll()] & \tt T,N,T,0 & 50 / 3,645,870 & 12s \\
3, 2, 2 & 100 / 312 & \tt [offer(0); poll()],[{\bf removeAll}([1,0])] & \tt T,0,T & 7,792 / 4,918,210 & 9s \\
4, 2, 2 & 300 / 1,788 & \tt [offer(0); {\bf removeLastOccurrence}(0)],[offer(0); poll()] & \tt T,F,T,0 & 90 / 3,975,720 & 95s \\
4, 2, 2 & 100 / 894 & \tt [offer(0); {\bf removeLast}()],[offer(0); poll()] & \tt T,E,T,0 & 18 / 3,435,730 & 23s \\
3, 2, 2 & 100 / 312 & \tt [offer(0); {\bf retainAll}([1,1])],[poll()] & \tt T,T,0 & 6,620 / 4,982,960 & 10s \\
4, 2, 2 & 100 / 894 & \tt [offer(1); poll(); offer(1)],[{\bf size}()] & \tt T,1,T,2 & 1 / 3,817,410 & 26s \\
4, 2, 2 & 100 / 894 & \tt [offer(1); poll(); offer(1)],[{\bf toArray}()] & \tt T,1,T,[1,1] & 68 / 2,156,940 & 9s \\
4, 2, 2 & 100 / 894 & \tt [poll(); offer(1)],[offer(0); {\bf toString}()] & \tt 0,T,T,[0,1] & 236 / 2,790,900 & 22s \\
\bottomrule
\end{tabular}
  \caption{Empirical results, Part 2.}
  \label{fig:experiments:2}
\end{table}

Each row of Tables~\ref{fig:experiments:1} and~\ref{fig:experiments:2} lists one
method in which our tool discovers an atomicity violation. The first column
lists the parameter values for enumeration: the number of invocations,
invocation sequences, and argument values. The second column lists the number of
harnesses tested versus the total number generated; we list multiples of $100$
since we invoke jcstress with $100$ harnesses per invocation. The next four
columns list the non-atomic harness reported, the non-atomic outcome observed
with that harness, the number of times jcstress observes that outcome, and the
total number of executions per harness, in one single second of testing. The
final column lists the total amount of wall-clock time passed before returning
the violating test harness. Note that harness testing is parallelized across
cores, while harness enumeration is computed sequentially on one core. The
outcome entries {\small\tt T}, {\small\tt F}, {\small\tt N}, {\small\tt E}
represent the values true, false, null, and an unspecified exception,
respectively. We represent lists and arrays with bracketed lists,
e.g.,~{\small\tt [0,1]}, and maps are represented with bracketed key-value
pairs, e.g.,~{\small\tt [0=1,1=1]}.

In most cases, testing exposes non-atomic outcomes with very low frequency. In
the most extreme case, jcstress witnesses a non-atomic outcome in
ConcurrentHashMap’s containsValue method only once in 3,993,110 executions,
i.e.,~with frequency~$2.50\times 10^{-7}$; the median frequency is $1.9\times
10^{-5}$, or roughly one in 5,000 executions. There are also degenerate cases,
like LinkedBlockingQueue’s removeAll method, in which over $10\%$ of executions
are non-atomic. Still, the fact that even the infrequent violations can be
observed in $1$ second of stress testing is what enables us to explore such a
large enumeration of harnesses. This is in stark contrast with the the Line-Up
tool~\cite{DBLP:conf/pldi/BurckhardtDMT10}, which performs systematic
exploration of concurrent schedules for a given harness, enumerating
linearizations per execution, spending an average of $31.5$ seconds per harness.
While their data does not allow us to determine the slowdown incurred by their
enumeration of linearizations per execution, since the number of executions per
harness is not given, we can conclude that our approach explores harnesses at
$31.5$ times the rate of Line-Up. Since Line-Up detects a violation in $23.7$
out of $100$ harnesses on average, their expected time for detecting a violation
is $31.5\text{s/h} \times 23.7\text{h} = 745.6\text{s}$. While this is
comparable to our average time of $352.5$s for detecting a violation (our
experiments are run on a $4$-core machine), we note that Line-Up requires
additional user insight in principle, since their users are required to specify
the arguments of each invocation used in a harness; our approach includes
invocation arguments in the discovery process, enabling the exposure of
violations which only occur with unexpected invocation-argument combinations.

Interestingly, while the number of invocations required to expose these
violations ranges from $3$–$6$, all are exposed with only $2$ invocation
sequences, and all but two are exposed with only $2$ argument values. The
exceptions are ConcurrentSkipListSet’s headSet($j$) and subSet($i,j$) methods
which return the subset of ordered elements smaller than $j$, excluding $j$, and
greater than $i$, including $i$, in the case of subSet. Since atomicity
violations appear only when at least $2$ elements $k_1, k_2$ are returned,
respectively $3$ elements $k_1, k_2, k_3$, thus $3$ and $4$ argument values are
required, i.e.,~to satisfy $k_1, k_2 < j$ and $i \le k_1, k_2, k_3 < j$.

\section{Discussion}
\label{sec:discussion}

In this work our goal is to expose atomicity violations in concurrent objects.
We believe this goal can be useful both for exposing implementation bugs, and
for exposing flaws or imprecisions in object specifications. While some of the
violations we have exposed are clear-cut cases of implementation bugs, many of
them might be categorized as either bugs or imprecise specifications. In some
cases, the lack of atomicity may even be expected according to a given
specification. In this section we discuss possible interpretations of our
results against the corresponding specifications\footnote{The Java Platform {\sc
se} 8 {\sc api} specification.
\url{https://docs.oracle.com/javase/8/docs/api/java/util/concurrent/package-summary.html}}.

The documentation for many bulk mutator methods like addAll, putAll, removeAll,
and retainAll, and read-only methods like size and toArray states fairly clearly
that atomicity is not intended. There are however several notable exceptions.
For instance, documentation for the bulk mutator methods in ArrayBlockingQueue,
PriorityBlockingQueue, and LinkedBlockingDeque do not mention atomicity. Still,
non-atomicity might be suspected from the fact many among these methods are
inherited from AbstractCollection and AbstractQueue, which rely on the
implementation of the iterator method, left abstract by AbstractCollection and
AbstractQueue, and ultimately declared to be “weakly consistent” by their
implementing classes. Other instances are the size and toArray methods of
LinkedBlockingQueue; their documentation provides no indication to whether
these operations may or may not be atomic.

In some cases where the documentation does indicate non-atomicity, our
non-atomicity violation still points to what is most certainly a bug. This is
the case with ConcurrentSkipListMap’s clear method whose violation, listed in
Figure~\ref{fig:tests:clear}, appears to break the object’s representation
invariant, possibly permanently.

We also find atomicity violations that contradict their documentation. For
instance, the clear method of LinkedBlockingQueue is explicitly stated to be
atomic. The case of the ConcurrentLinkedDeque is also interesting because the
documentation claims that “concurrent insertion, removal, and access operations
execute safely across multiple threads,” although we find that the addFirst,
pollLast, peekLast, and removeLastOccurrence methods are certainly not atomic.
The same holds for the pollFirst and pollLast methods of ConcurrentSkipListSet.

Finally, there are several methods for which the documentation contains no
reference whatsoever to atomicity. For instance, this is the case with the
remove and overloaded contains(Object) methods of LinkedBlockingQueue, and the
tailSet, headSet, and subSet methods of ConcurrentSkipListSet.

\section{Related Work}
\label{sec:related}

Given the acknowledged difficulty of concurrent programming, there are many
lines of research which aim to enable the construction of reliable concurrent
software. Much of that research focuses on techniques for proving correctness of
an implementation, e.g.,~automatically verifying programmer-supplied inductive
invariants~\cite{DBLP:journals/acta/OwickiG76}, and, e.g.,~automatically
constructing and verifying inductive invariants,
e.g,.~\cite{DBLP:conf/popl/CousotC77}. In contrast, our work focuses on
falsifying program properties. While our approach cannot prove the correctness
of a given implementation, any reported violations are guaranteed to be actual
violations. The other key advantages of our approach are that (a)~it does not
require annotation of, or even knowledge of, the implementation code, and
(b)~since our output is simple test harnesses, e.g.,~rather than complicated
program invariants, it is immediately familiar and useful to programmers.

While our approach to testing a given harnesses is \emph{stress testing},
i.e.,~subjecting a given harness to high loads and parallelism, often millions
of executions per second, there are a few other notable approaches. Systematic
concurrency testing tools assert control over the schedule in which threads’
instructions are executed, and systematically explore each possible thread
schedule, or at least a limited set of representative
schedules~\cite{DBLP:conf/pldi/MusuvathiQ07}. In principle that approach is
guaranteed to find any violation in a given test harness given enough time; in
practice, the combinatorial explosion of possible thread interleavings prohibits
this exploration to a limited set of representatives, e.g.,~small but prescribed
perturbations from some deterministic scheduling. Recent work has shown
approaches to cut down on this combinatorial explosion by having the programmer
express which points of the program require rigorous scheduling exploration,
enumerating alternative schedules only around those
points~\cite{DBLP:conf/pldi/ElmasBNS13}. Probabilistic concurrent
testing~\cite{DBLP:conf/asplos/BurckhardtKMN10} explores a broader sampling of
schedules using randomness, yielding probabilistic guarantees. Compared to those
approaches, the simple stress testing we use provides no coverage guarantees; it
has the advantage of being very fast since it avoids possibly-costly mechanisms
for controlling scheduling, e.g.,~lock-based synchronization. All of the above
approaches assume a fixed test harness.

Concolic testing~\cite{DBLP:conf/pldi/GodefroidKS05, DBLP:conf/sigsoft/SenMA05}
partially addresses the problem of test-harness generation by repeatedly
generating program inputs, e.g.,~the arguments to invocations in the test
harness, using prior test executions. During each execution, that approach
records the results of branching instructions, along with constraints relating
branch decisions to program input values; inputs for the next execution are
generated by constraint solving for an execution that makes different branch
decisions. In practice, this approach would be effective at generating
invocation arguments for a given test harness with unspecified arguments, but
likely not at generating the relevant sequences of invocations. While concolic
testing was initially developed for sequential programs, it has been extended to
assert control over scheduling decisions as well as branch decisions, invoking
similar comparisons with those mentioned above between stress testing and
systematic and probabilistic concurrency testing.

Some works have investigated the generation of entire test harnesses rather than
only the invocation arguments of a given template harness. These works are
typically focused on one particular domain of program properties. For instance,
some works automatically generate functional-conformance tests from given formal
specifications~\cite{DBLP:conf/memocode/GongWSDWM12}, small litmus-test programs
which exhibit non-sequentially-consistent memory
behaviors~\cite{DBLP:conf/tacas/AlglaveMSS11, DBLP:conf/dac/Mador-HaimAM11}, or
small programs which exhibit bugs in compiler
optimizations~\cite{DBLP:conf/pldi/ChenGZWFER13}. The most-closely related work
to ours also generates concurrent data structure test
harnesses~\cite{DBLP:conf/pldi/BurckhardtDMT10}; given a set of $I$ invocations
with fixed arguments, and parameters $T, S, N$ respectively for the number $T$
of invocation sequences (threads), the number $S$ of invocations per sequence,
and the total number $N$ of generated harnesses, they generate $N$ combinations
of $T$ length-$S$ invocation sequences using invocations $I$. Compared to our
harness enumeration, theirs is less complete in principle, since it does not
consider alternative argument values, nor ordering between invocation sequences,
and limits the number of combinations. Our characterization of core methods
enables us to explore more interesting combinations of a single non-core method
invocation with core method invocations, and our efficient atomicity check
enables us to explore harnesses at $31.5$ times their rate.

Our notion of atomicity is the embodiment of Liskov’s substitution
principle~\cite{DBLP:journals/toplas/LiskovW94} known as
\emph{linearizability}~\cite{DBLP:journals/toplas/HerlihyW90} which is now known
to correspond exactly to the client-centric notion observational
refinement~\cite{DBLP:journals/tcs/FilipovicORY10}; intuitively, this states
that any behavior exhibited by a client of a given concurrent object could have
also been exhibited by the same client were it using a truly-atomic
implementation, i.e.,~where operations’ steps are not allowed to overlap. While
this notion is often the most appropriate for shared-memory concurrent objects,
there are other interesting notions of consistency:
serializability~\cite{DBLP:journals/jacm/Papadimitriou79b} captures the
atomicity of transactional sequences, e.g.,~sequences of database operations;
eventual consistency~\cite{DBLP:conf/sosp/TerryTPDSH95,
DBLP:journals/ftpl/Burckhardt14} and causal
consistency~\cite{DBLP:journals/cacm/Lamport78} capture weaker notions of
consistency for, e.g.,~distributed key-value stores, where the cost of
synchronization required to ensure linearizability is deemed too expensive.

Testing such atomicity properties generally requires more instrumentation than
the simple annotation of program assertions;
\citet{DBLP:journals/entcs/TasiranQ05} survey techniques for the runtime
verification of atomicity. Initial approaches to testing recorded the
happens-before order of executed operations, and enumerated possible
linearizations while verifying that some linearization corresponds to an atomic
behavior~\cite{DBLP:journals/jpdc/WingG93, DBLP:conf/pldi/BurckhardtDMT10,
DBLP:conf/asplos/BurnimNS11, DBLP:journals/tse/ZhangCW15}; these approaches are
intractable in the length of executions. This basic approach has been optimized
for set-type concurrent data structures by exploiting the fact that the
operations corresponding to distinct keys can be separately linearized, reducing
the amount of combinatorial explosion in
linearization~\cite{DBLP:journals/concurrency/Lowe17, DBLP:conf/forte/HornK15a}.
Alternative approaches have pursued tractability by recording an approximation
of the happens-before order of executed
operations~\cite{DBLP:conf/popl/BouajjaniEEH15}, and by lazily applying
type-specific rules to ordering recorded
operations~\cite{DBLP:conf/pldi/EmmiEH15}; while tractable, these approaches are
not guaranteed to catch a given atomicity violation. In contrast to all of the
aforementioned works, ours can be seen as an optimization to linearizability
checking for a fixed loop-free test harness: the invocations of such a harness
can be indexed statically, which enables our notion of result-vector outcomes;
the expensive enumeration can then be replaced by a simple comparison of outcome
with the precomputed atomic outcomes, which are typically much fewer than
linearizations.

A final point of comparison is with works on transactional
boosting~\cite{DBLP:conf/ppopp/HerlihyK08} and linearizability checking for
transactions of linearizable operations~\cite{DBLP:conf/oopsla/ShachamBASVY11}.
These methods reason about atomicity of core-method compositions using
commutativity arguments in the composed-method implementations. Such techniques
would be applicable to checking atomicity of many non-core methods such as
addAll, containsAll, \ldots, which could in principle be implemented by
compositions of core methods. In contrast to these techniques, our uncovers
atomicity violations, exhibiting non-atomic test harnesses, without any
knowledge of the implementation code.

\section{Conclusion}
\label{sec:conclusion}

Our results demonstrate an effective method for exposing atomicity violations in
concurrent objects, and a great number of violations in the most widely-used
software development kit: The {\sc jdk}. While we have not yet contacted the
maintainers of these implementations to report our findings in respect of
double-blind submission guidelines, we do soon intend to, i.e.,~via the
Concurrency Interest mailing list. As discussed in Section~\ref{sec:discussion},
while we have identified some clear-cut implementation bugs, e.g.,~in
ConcurrentSkipListMap’s clear method, we suspect that many of these atomicity
violations may not be considered bugs, rather implicitly-allowed behavior of
imprecise specifications. In any case, we believe that our findings would be of
interest to both the maintainers and users of {\sc jdk} concurrent data
structures. Additionally, our results may also be of interest to users and
maintainers of other concurrent object libraries, e.g.,~in other languages,
since our approach is fairly agnostic to the language and library in question:
see Section~\ref{sec:experiments} for a short discussion on limitations and
Java-language dependencies.


\bibliography{dblp}


\end{document}